\documentclass[aps,prl,twocolumn,showpacs]{revtex4}  

\usepackage{graphicx}  
\usepackage{amsmath}
\usepackage[utf8]{inputenc}   
\usepackage[T1]{fontenc}      
\usepackage{textcomp}
\pdfoutput=1

\begin{document}
\title{Exploiting spatiotemporal degrees of freedom for far field subwavelength focusing using time reversal in fractals}
\author{Matthieu Dupré, Fabrice Lemoult, Mathias Fink, Geoffroy Lerosey}
\email[]{geoffroy.lerosey@espci.fr}
\affiliation{Institut Langevin, ESPCI Paris \& CNRS UMR 7587, 1 rue Jussieu, 75005 Paris, France}
\date{\today}

\begin{abstract}
Materials which possess a high local density of states varying at a subwavelength scale theoretically permit to focus waves onto focal spots much smaller than the free space wavelength. To do so metamaterials -manmade composite media exhibiting properties not available in nature- are usually considered. However this approach is limited to narrow bandwidths due to their resonant nature. Here, we prove that it is possible to use a fractal resonator alongside time reversal to focus microwaves onto $\lambda/15$ subwavelength focal spots from the far field, on extremely wide bandwidths. We first numerically prove that this approach can be realized using a multiple channel time reversal mirror, that utilizes all the degrees of freedom offered by the fractal resonator. Then we experimentally demonstrate that this approach can be drastically simplified by coupling the fractal resonator to a complex medium, here a cavity, that efficiently converts its spatial degrees of freedom into temporal ones. This allows to achieve deep subwavelength focusing of microwaves using a single channel time reversal. Our method can be generalized to other systems coupling complex media and fractal resonators.
\end{abstract}

\maketitle

Controlling the propagation of waves in complex media is of fundamental interest in a wide range of research fields from imaging in biological tissues to geophysics and telecommunications. In the past, it was shown that wave-front shaping techniques permit to image and focus waves in and through multiple scattering or reverberating media. Such methods can experimentally vary between the different fields: Time reversal is used in microwaves and acoustics~\cite{fink1992,fink2008,derode2001,lerosey2004}, whereas  phase conjugation~\cite{steel2010,papadopoulos2012}, spatial light modulators~\cite{vellekoop2010,katz2011,park2013}, or photoacoustics~\cite{lai2015} are used in optics. However, those methods rely on the same basis: They make use of spatial or temporal degrees of freedom~\cite{draeger1997,vellekoop2007,mosk2012} or both~\cite{derode2001,lerosey2006,lemoult2009,mccabe2011,katz2011,aulbach2011,andreoli2015} to focus waves in complex media. So called spatial degrees of freedom correspond to the number of independent collected modes at a fixed frequency, whereas the temporal degrees of freedom refer to the number of uncorrelated modes within a given bandwidth.

Focusing waves in complex media amounts to coherently add such modes either at a given location with wave-front shaping or at a given time and location with time reversal.  In the vicinity of the focal spot the modes interfere constructively in a spatiotemporal window, and destructively out of it. The dimensions of such coherence window are fixed by the correlation time, given by the inverse of the bandwidth of the modes supported by the medium, and correlation length of the field, given by the highest spatial frequency of the eigenmodes~\cite{derode2001}.

Reducing the size of the focal spot down to subwavelength dimensions is of prime importance for bio-imaging or nanolithography applications. This requires media with a LDOS varying at the subwavelength scale which can be achieved with subwavelength varying random media~\cite{li2008,park2013,park2014,caze2013,gjonaj2013,lanoy2015}. Yet such approach is very limited in resolution in optics because of the small values of the electric permittivity. Another approach consists in using metamaterials that can efficiently manipulate the evanescent waves~\cite{veselago1968,pendry2000,liu2007,li2011,lu2012,lemoult2010,lemoult2011,lemoult2011-1,lemoult2011-2,lemoult2012}. However, the resonant nature of such materials restricts such approach to narrow bandwidths, which limits the number of degrees of freedom that can be harnessed. Moreover, dissipation not only equally restrains the number of degrees of freedom but also the field of view of metamaterial based lenses.

In this Letter, we prove in the microwave domain that it is possible to focus waves at deep subwavelength scales and on very wide bandwidths, using a fractal resonator. To do so we first use a one channel time reversal mirror~\cite{draeger1997} and a Hilbert fractal resonator of order $6$. Yet, we prove that such approach results in a very low focusing quality owing to the poorly resonant nature of the fractal. Hence we simulate a multiple channel time reversal in order to increase the number of exploited spatial degrees of freedom, and show that it allows deep subwavelength focusing of microwaves with very low residual sidelobes. We finally propose to simplify drastically this approach by coupling the fractal to a complex medium, that converts these spatial degrees of freedom into temporal ones. Henceforth we use a very simple experimental apparatus consisting of a reverberating cavity opened by a fractal resonator, and demonstrate experimentally focal spots as small as $\lambda/15$, obtained with one channel far field time reversal. 

Fractals~\cite{mandelbrot1984} are geometrical objects with a Hausdorff dimension~\cite{hausdorff1918} which is different from their topological dimension. They also possess self-similarity and scale invariant  properties: A phenomenon occurring at a given scale also occurs at many other ones. Hence, a fractal resonator exhibits in a wide bandwidth many log-periodic resonances ~\cite{engheta2006,wen2002,tanese2014}. Those scale invariant properties are widely used in physics and engineering to design metamaterials~\cite{engheta2006,hou2008,wen2005,xu2013,volpe2011,huang2010}, wide band antennas~\cite{best2002,werner2003}, filters~\cite{wen2002,barra2005,wen2003}, cavities and diffusers~\cite{jordan1983,dantonio1998} in optics, microwaves and acoustics.

We decide to use a planar fractal in order to manipulate a convenient flat lens. Moreover focusing waves wherever in its near field also requires the fractal to be as homogeneous as possible. Hence, we choose to work with the Hilbert curve (Fig.~1(a) presents a Hilbert curve of order 4). The Hilbert curve is one dimensional but fills a two dimensional plane: It is a one dimensional object with a Hausdorff dimension of two. The total length of the $n$th fractal order is $L_n=l_0 (2^n-1/2^n )$ for a footprint of $l_0^2$: Its total length $L_n$ is much larger than its apparent length $l_0$. Therefore a metallic Hilbert fractal is nothing else than a folded wire, and its fundamental mode schematically occurs at a wavelength $\lambda_n\approx L_n/2\approx l_0 2^{n-2}$. In other words, a Hilbert resonator possesses a footprint much smaller than the wavelength at resonance and constitutes a subwavelength resonator. Here, we use this property to set a high number of subwavelength modes in our bandwidth: the higher the fractal order, the higher the LDOS and the more subwavelength the modes. Dissipation is also limited as we work in transmission on a flat resonator with waves impinging on the transverse size of the fractal, contrary to previous works~\cite{lemoult2010,lemoult2011-1}. Hence the Hilbert curve provides three main benefits compared to other resonators: It fills a two-dimensional plane, it exhibits many resonances, and such resonances are subwavelength and occur in a very wide bandwidth.

\begin{figure}[t]
	\begin{center}
	\includegraphics[width=1\columnwidth]{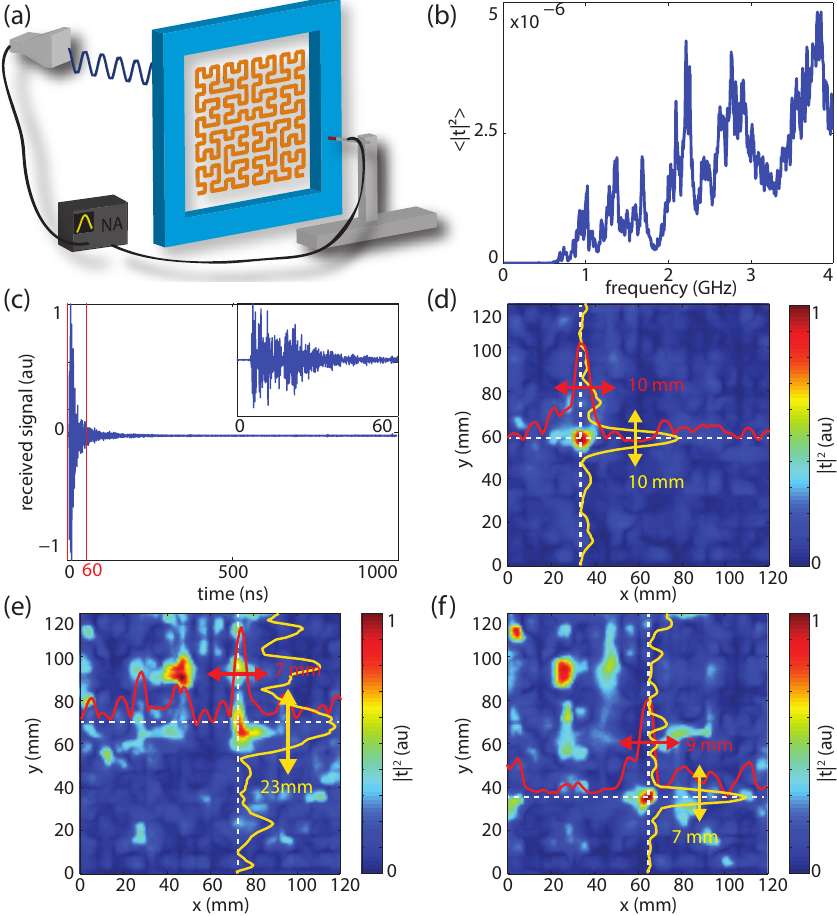}
	\caption{Focusing with a Hilbert resonator. (a)~Experimental set-up: A network analyzer measures the transmission of a metallic Hilbert curve inserted in a metallic screen  between a horn antenna and a near field probe. . (b)~Transmission spectrum averaged on the positions. (c)~Typical transmission signal at $\mathbf{r_1}$=(34~mm,59~mm). (d-f)~Time Reversal focusing (maximum over time of the energy) at positions $\mathbf{r_1}$, $\mathbf{r_2}=(73,70)$, and $\mathbf{r_3}=(66,35)$. Red and yellow curves give the profiles on the dashed white lines which intersections picture the focal spots.}
	\label{fig1}
	\end{center}
\end{figure}

We start by measuring the transmission through a Hilbert curve of order 6 from the far field. The set-up is shown on Fig.~1(a): A network analyzer measures the transmission through a metallic fractal. The latter  is made of copper and is printed with a PCB lithography technique on a dielectric substrate. Its footprint is 120~mm by 120~mm wide for a total length of 7.7~m. As this footprint corresponds to the wavelength at 2~GHz, the resonator supports a very large number of subwavelength resonances that permit the waves to go through, although the holes in the metal are very small (around a few millimeters) compared to the wavelength~\cite{wen2005}. On one side a horn antenna emits microwaves within the 1.5~GHz to 3~GHz bandwidth from the far field at 1.5~m of the fractal, and the network analyzer measures the received voltage on a probe placed on the other side in the near field of the fractal. The latter is placed at 0.5~mm of the fractal on a two dimensional translation stage which scans the plane in a 120 by 120~mm$^2$ area.

The network analyzer measures a transmission spectrum for all probe positions, from which we obtain the transient Green's function of the medium with an inverse Fourier transform. Fig.~1(b) presents the transmission spectrum averaged on the positions. Fig.~1(c) presents the measured time signal at a given position of the near field probe: After a 5~ns delay corresponding to the time of flight for the 1.5~m distance between the two antennas, a signal is received. The resonating nature of the fractal lengthens the initial 0.67~ns pulse, and we measure a longer coda that decreases with a characteristic time of 15~ns. Within the 1.5~GHz to 3~GHz bandwidth, this signal provides 12 temporal degrees of freedom~\cite{supplemental}, that we can use to focus the wave field at a given time and position, $\mathbf{r_0}$. We use here a broad band approach to synchronize the subwavelength modes: time reversal. Experimentally, the time signal recorded at the position $\mathbf{r_0}$, would be time reversed and sent back from the horn antenna, and the near field probe would measure the field received at any position $\mathbf{r}$. We do so numerically as this amounts to compute the cross-correlation of the signal measured at position $\mathbf{r_0}$ with the signal measured at position $\mathbf{r}$. Fig.~2(d) presents the maximum value over time of the energy of the computed field. A subwavelength spot of size $\lambda/10$ (where $\lambda$ is the central wavelength) is obtained at the focal position $\mathbf{r_0}$ with a single antenna emitting from the far field. However for other targeted locations (Fig.~2(e,f)), there are supplementary illuminated hot spots, even far away from the target, in addition to a high background. Even if subwavelength focal spots are obtained, the low quality background and the additional hot spots decrease the focusing quality with a low signal to noise ratio (SNR)~\cite{supplemental}.
\begin{figure}
	\begin{center}
	\includegraphics[width=1\columnwidth]{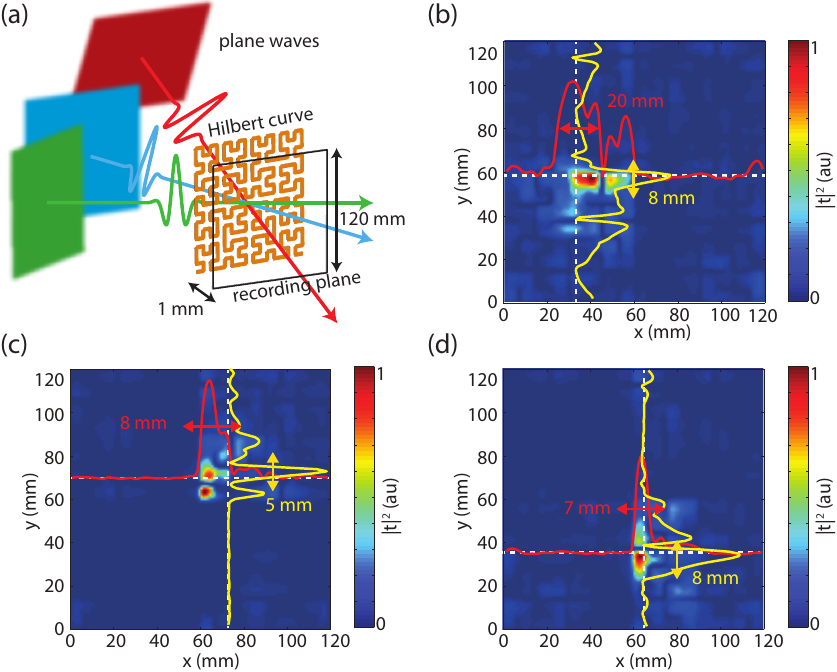}
	\caption{Simulations: focusing with an order 6 Hilbert fractal and 10 spatial degrees of freedom. (a)~Simulated set-up. The Hilbert curve is illuminated by 90 plane waves (45 angles of incidence and two orthogonal polarizations) of Gaussian pulses, three of them are represented in red, blue and green. The electric field is recorded in plane parallel to the fractal at a distance of 1~mm. (b-d)~Time maximum value of the energy when focusing at positions $\mathbf{r_1}$,  $\mathbf{r_2}$ and $\mathbf{r_3}$.}
	\label{fig2}
	\end{center}
\end{figure}

In order to improve the focusing quality and the SNR, a solution is to increase the number of spatial degrees of freedom, \textit{i.e.} the number of source antennas. Therefore we decide to illuminate the fractal with 90 different plane waves (45 different incident angles and 2 orthogonal polarizations~\cite{supplemental}). We opt for a numerical study (using CST Microwave studio) as such a procedure would be experimentally cumbersome. First, we simulate the fields excited by each of those 90 plane waves. However those plane waves produce only 10 uncorrelated fields on the fractal~\cite{supplemental}, meaning that the number of spatial degrees of freedom is $N_s=10$ instead of the expected 90. We evaluate the number of temporal degrees of freedom to $N_t=22$~\cite{supplemental}. The latter is higher than in previous measurements, but of the same order of magnitude. Hence the total number of spatiotemporal degrees of freedom is $N_{tot}=N_s\times N_t=220$. We then use time reversal to synchronize those degrees of freedom at a given time and at position $\mathbf{r_0}$. To do so, the signals previously measured at $\mathbf{r_0}$ for every plane wave are time reversed and sent back through their corresponding plane wave. We record the sum of the interfering fields in a plane at 1~mm of the fractal. Three focal spots displayed on Fig.~2 exhibit very sub-wavelength dimensions (around $\lambda/10$), and a very low background. Those results prove that increasing the number of spatial degrees of freedom by an order of magnitude efficiently improves the focusing capabilities of the fractal.

\begin{figure}
	\begin{center}
	\includegraphics[width=1\columnwidth]{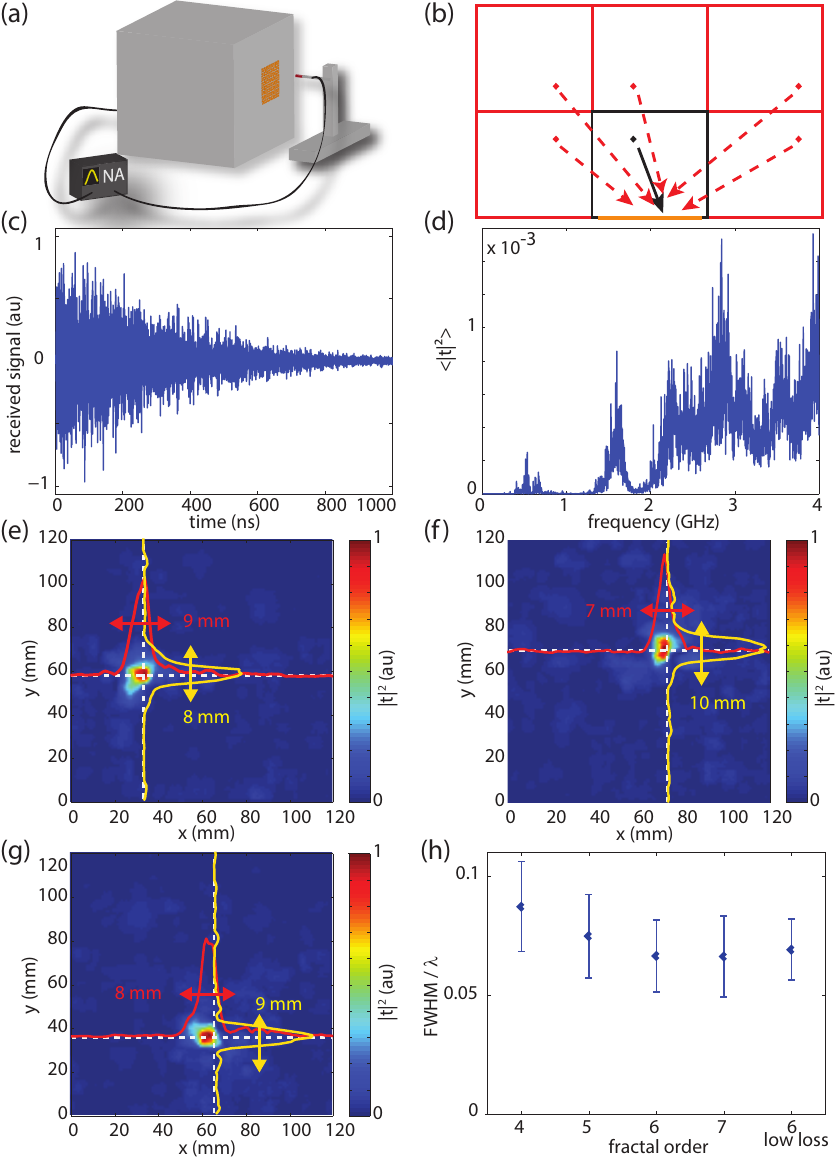}
	\caption{Focusing with a Hilbert fractal and a cavity. (a)~Experimental set-up: A network analyzer measures the transmission of a metallic Hilbert fractal inserted in the wall of a high Q cavity,  between an inside isotropic antenna and a near field probe. (b)~Schematic view of the increase of degrees of freedom: black: real cavity, red: image sources and cavities. (c)~Time signal in transmission at $\mathbf{r_1}$. (d)~Transmission spectrum averaged on the positions. (e-g)~Time reversal focusing with a cavity at positions $\mathbf{r_1}$,  $\mathbf{r_2}$ and $\mathbf{r_3}$. (h)~Average size (100 random positions), and standard deviation (error bars) of the focal spots obtained with fractals of order  4 to 7 with usual dielectric substrate and with an order 6 made with a low loss substrate.}
	\label{fig3}
	\end{center}
\end{figure}

However using such a number of transient sources is experimentally challenging. As a consequence, we propose a very simple experimental set-up shown on Fig.~3(a). We use a steel commercial cavity of 1~m$^3$ volume, opened on one side by the fractal (Fig.~3(a)). The Q factor is about 1800. An isotropic antenna is placed inside the cavity to replace the horn antenna of Fig.~1(a). As schemed on Fig.~3(b), the cavity creates fictive sources as mirror images of the real ones, and which provide additional degrees of freedom (\textit{i.e.} incident wave vectors). As there is only one real source, the number of spatial degrees of freedom is 1. However, the fictive sources provide additional temporal degrees of freedom by increasing the time of flight from the emitters to the fractal. Hence the temporal signal measured is much longer with a cavity than without: The number of temporal degrees of freedom has been increased. Indeed, as shown on the Fig.~3(c), the measured signal now attenuates more slowly in a characteristic time of 300~ns: The coda lasts 20 times longer. We note that the envelope of the transmission spectrum with a cavity is quite different than without it, as the horn antenna has been replaced by an isotropic WI-FI antenna, which operates in a narrower bandwidth. Therefore, the total number of spatiotemporal degrees of freedom is $N_{tot}=150$ \cite{supplemental}, a bit lower than in simulations, but of the same order of magnitude, and much higher than the number of degrees of freedom obtained in free space with only one antenna. This enhancement is directly translated into the spectral domain: The initial modes are now 20 times better resolved than with only one antenna, as we see on the averaged transmission (Fig.~3(d)). 

Now that we have experimentally increased by an order of magnitude the number of degrees of freedom of the fractal resonator, we can focus waves with one-channel time reversal, as we did in the first experiment. Fig.~3(e-g) show the results with the maximum value over time of the energy at each position of the measurement plane. We see subwavelength focal spots at the desired places, with full widths at half maximum around 8~mm, less than $\lambda /15$. Contrary to the case without a cavity on Fig.~2(d), there is no side lobe and the background is very low with a high SNR. The results are equivalent to the simulation results with multiple illuminations, but thanks to the cavity, we obtained them with a single emitter.

In order to study the impact of the fractal order, we run similar experiments for orders ranging from 4 to 7. The average focal widths for each order are shown in Fig.~3(h). The sizes of the focal spots decrease with the fractal order, owing to the fact that the higher the order, the smaller the coherence length of the resonator. However, this phenomenon saturates for the highest measured fractal orders. In such cases, the modes have a very high spectral density, and cannot be resolved anymore even by our cavity, which leads to the saturation of the number of temporal degrees of freedom. A Hilbert fractal of order 6 printed on a low loss dielectric substrate (NELTEC NH9338ST, tangent loss $\delta=3~10^{-3}$), does not show significant differences compared to the FR4 substrate ($\tan \delta=3~10^{-2}$), underlying the fact that the modes of the Hilbert fractal of order 6 are just resolved by the cavity which limits the focusing for higher fractal orders.

\begin{figure}
	\begin{center}
	\includegraphics[width=1\columnwidth]{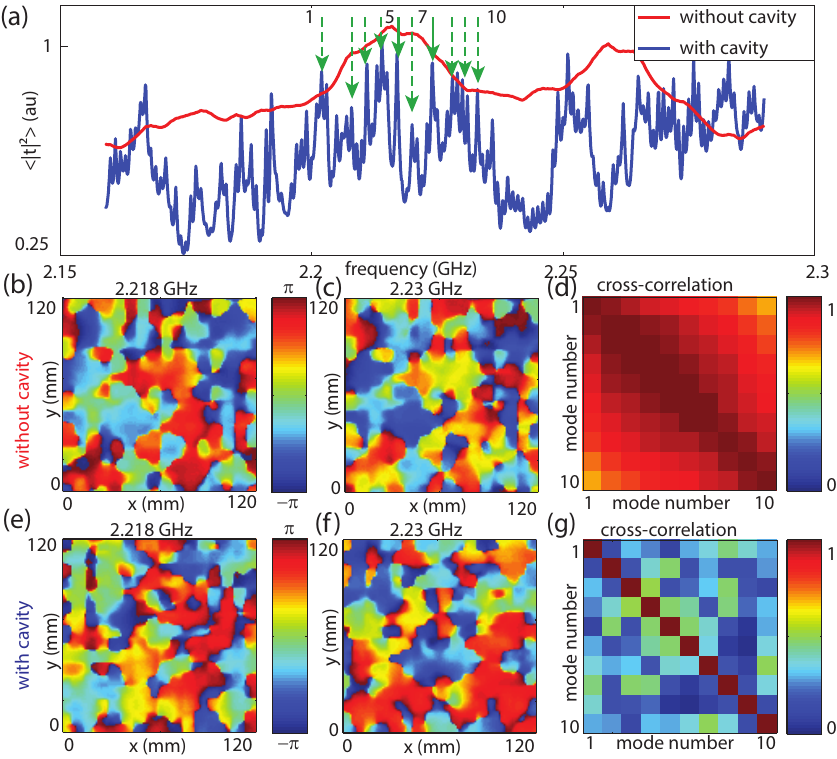}
	\caption{Comparing the spatial coherence. (a)~Measured transmission spectrum of a Hilbert fractal of order 6 without and with a cavity between 2.18 and 2.28~GHz. The green dashed arrows refer to the modes picked up for (d) and (g). (b)~and (c) Phase of the modes 1 and 10 at 2.18 and 2.23~GHz without a cavity. (d)~Spatial cross-correlations of the modes measured without a cavity. (e-g)~Same as (b-d) with a cavity.}
	\label{fig4}
	\end{center}
\end{figure}

The last point to clarify is the role of the cavity. Namely, how can this complex medium exploit optimally the degrees of freedom of the fractal resonator ? To illustrate this point, we compare the modes measured in a narrow bandwidth without and with a cavity. Fig.~4(a) presents the considered bandwidth of 2.18 to 2.28~GHz, with a smooth transmission spectrum without a cavity, and well resolved modes for the transmission with a cavity. In this bandwidth, we select 10 frequencies close to each other (green arrows of Fig.~4(a)). Fig.~4(b) and (c) show the phase of the transmission of the fields at two among of those frequencies (2.218~GHz and 2.23~GHz) measured without a cavity : They are almost identical. Indeed, the cross-correlation coefficients (Fig.~4(d)) of those ten modes are very high (above 0.7). Hence the electromagnetic fields are the same at any of these frequencies: There is only one temporal degree of freedom in this bandwidth. On the contrary, the fields obtained with the cavity at the first and last frequencies are very different (Fig.~4(e) and (f)) as confirmed by the very low cross-correlation coefficients of the ten modes inferior to 0.5. Hence, with a cavity, all those modes are quasi-uncorrelated: Any of them  provide a temporal degree of freedom for the focusing. The cavity provides spectrally distinguishable illuminations on the fractal resonator within one resonance of the fractal. This permits to replace the multiple illuminations thanks to the complexity of the cavity: a single spatial degree of freedom (one source) can provide multiple temporal degrees of freedom. 

In this Letter, we have achieved sub-wavelength focusing down to $\lambda/15$ from the far field using a fractal resonator and time reversal in the microwave domain. Playing with the spatial and temporal degrees of freedom, we have shown that fractals are good candidate to achieve subwavelength resolution and proved that by adding a reverberating medium, multiple illuminations can be replaced by a single one. We also illustrated the impact of the fractal order: a fractal resonator possesses a very high number of low frequency modes. The majority of those subwavelength modes cannot be excited by a single source. To reveal and make use of such modes, it is necessary to increase the number of degrees of freedom. We believe that such approach, that is coupling a complex medium with a fractal resonator can be generalized to other domains. For instance in optics, one could use a random medium to increase the number of degrees of freedom and to resolve the modes of a metallic film at the percolation threshold~\cite{krachmalnicoff2010}, which is fractal.

M.D. acknowledges fundings from French ``Ministère de la Défense, Direction Générale de l’Armement''. This work is supported by LABEX WIFI (Laboratory of Excellence within the French Program ``Investments for the Future'') under references ANR-10-LABX-24 and ANR-10-IDEX-0001-02 PSL* and by Agence Nationale de la Recherche under reference ANR-13-JS09-0001-01.

\bibliography{biblio}

\end{document}